\RequirePackage{lineno}

\documentclass[aps,prd,twocolumn,groupedaddress,showpacs,floatfix,letterpaper]{revtex4}
\usepackage{verbatim}

\usepackage{graphicx}
\usepackage{multirow}
\usepackage{amsfonts}
\usepackage{amsmath}
\usepackage{amssymb}
\usepackage[hidelinks=true]{hyperref}
\usepackage{url}
\usepackage{array}
\usepackage{placeins}
\usepackage{bm}
\usepackage{xspace}
\usepackage{booktabs}           
\usepackage[normalem]{ulem}
\usepackage{units}
\usepackage{float}
\usepackage{array}
\usepackage{lineno}
\bibpunct{[}{]}{,}{n}{}{}

\def\gev {\ensuremath {\rm \,GeV}\xspace}
\def\mev {\ensuremath {\rm \,MeV}\xspace}

\xspace 
\xspace
\def\gevsq {\ensuremath {\rm \,GeV}^2\xspace}

\def\den {\ensuremath {\rm \,g}\!/\!cm^3}\xspace
\xspace
\def\gev {\ensuremath {\rm \,GeV}\xspace}

\def\gevc {\ensuremath {\rm \,GeV}\!/\!c\xspace}

\def\mev {\ensuremath {\rm \,MeV}\xspace}

\def\cm {\ensuremath {\rm \,cm}\xspace}

\def\m {\ensuremath {\rm \,m}\xspace}

\def\cmsq {\ensuremath {\rm \,cm}^{2}\xspace}
\def\ch2{\ensuremath {\rm \,CH_{2}}\xspace}

\begin{document}
\newcommand{\makespace}{\vspace{3 mm}}
\newcommand{\DP}{\displaystyle}

\title{Measurement of the Antineutrino Neutral-Current Elastic Differential Cross Section}

\date{\today}

\author{
        A.~A. Aguilar-Arevalo$^{12}$, 
        B.~C.~Brown$^{6}$, L.~Bugel$^{11}$,
	G.~Cheng$^{5}$, E.~D.~Church$^{15}$, J.~M.~Conrad$^{11}$,
	R.~Dharmapalan$^{1}$, 
	Z.~Djurcic$^{2}$, D.~A.~Finley$^{6}$, R.~Ford$^{6}$,
        F.~G.~Garcia$^{6}$, G.~T.~Garvey$^{9}$, 
        J.~Grange$^{7}$,
        W.~Huelsnitz$^{9}$, C.~Ignarra$^{11}$, R.~Imlay$^{10}$,
        R.~A. ~Johnson$^{3}$, G.~Karagiorgi$^{5}$, T.~Katori$^{11}$\footnote{Present address: Queen Mary University of London, London E1 4NS, United Kingdom.},
        T.~Kobilarcik$^{6}$, 
        W.~C.~Louis$^{9}$, C.~Mariani$^{16}$, W.~Marsh$^{6}$,
        G.~B.~Mills$^{9}$,
	J.~Mirabal$^{9}$,
        C.~D.~Moore$^{6}$, J.~Mousseau$^{7}$, 
        P.~Nienaber$^{14}$, 
        B.~Osmanov$^{7}$, Z.~Pavlovic$^{9}$, D.~Perevalov$^{6}$,
        C.~C.~Polly$^{6}$, H.~Ray$^{7}$, B.~P.~Roe$^{13}$,
        A.~D.~Russell$^{6}$, 
	M.~H.~Shaevitz$^{5}$, 
        J.~Spitz$^{11}$, I.~Stancu$^{1}$, 
        R.~Tayloe$^{8}$, R.~G.~Van~de~Water$^{9}$, M.~O.~Wascko$^{17}$,
        D.~H.~White$^{9}$, D.~A.~Wickremasinghe$^{3}$, G.~P.~Zeller$^{6}$,
        E.~D.~Zimmerman$^{4}$ \\
\smallskip
(MiniBooNE Collaboration)
\smallskip
}
\smallskip
\smallskip
\affiliation{
$^1$University of Alabama; Tuscaloosa, AL 35487 \\
$^2$Argonne National Laboratory; Argonne, IL 60439 \\
$^3$University of Cincinnati; Cincinnati, OH 45221\\
$^4$University of Colorado; Boulder, CO 80309 \\
$^5$Columbia University; New York, NY 10027 \\
$^6$Fermi National Accelerator Laboratory; Batavia, IL 60510 \\
$^7$University of Florida; Gainesville, FL 32611 \\
$^8$Indiana University; Bloomington, IN 47405 \\
$^9$Los Alamos National Laboratory; Los Alamos, NM 87545 \\
$^{10}$Louisiana State University; Baton Rouge, LA 70803 \\
$^{11}$Massachusetts Institute of Technology; Cambridge, MA 02139 \\
$^{12}$Instituto de Ciencias Nucleares, Universidad Nacional Aut\'onoma de M\'exico, D.F. 04510, M\'exico \\
$^{13}$University of Michigan; Ann Arbor, MI 48109 \\
$^{14}$Saint Mary's University of Minnesota; Winona, MN 55987 \\
$^{15}$Yale University; New Haven, CT 06520\\
$^{16}$Center for Neutrino Physics, Virginia Tech; Blacksburg, VA 24061\\
$^{17}$Imperial College London, London SW7 2AZ, United Kingdom\\
}

\begin{abstract}
We report the measurement of the flux-averaged antineutrino neutral current elastic scattering cross section ($d\sigma_{\bar \nu N \rightarrow \bar \nu N}/dQ^{2}$) on CH$_{2}$ by the MiniBooNE experiment using the largest sample of antineutrino neutral current elastic candidate events ever collected.
The ratio of the antineutrino to neutrino neutral current elastic scattering cross sections and a ratio of antineutrino neutral current elastic to antineutrino charged current quasi elastic cross section is also presented. 
\vspace{1.0in}
\end{abstract}

\maketitle
\pagebreak[4]


\section{Introduction.}
One of the simplest weak neutral current interactions is the elastic scattering of a neutrino from a nucleon (NCE).
This process is sensitive to both isoscalar and isovector weak currents carried by the nucleon whereas charge current quasi-elastic (CCQE) scattering is sensitive to only the isovector current.
Both NCE and CCQE neutrino interactions are important for accelerator-based neutrino oscillation experiments, and to date, very few measurements in the GeV regime have been made, particularly with antineutrinos~\cite{xsec_bible}.

Recent measurements of neutrino-nucleus CCQE scattering on $^{12}$C show an enhanced cross section~\cite{MB_CCQE} relative to the prediction from impulse approximation calculations, such as the relativistic Fermi gas (RFG) model of the nucleus~\cite{Smith-Moniz}. 
The enhancement is likely to arise from nucleon-nucleon correlations absent in the RFG model and NCE scattering provides a complementary channel to further examine the nuclear effects common to both CCQE and NCE neutrino-nucleon scattering.

The Mini Booster Neutrino Experiment (MiniBooNE) has previously reported high-statistics measurements of neutrino CCQE ($\nu$CCQE) and neutrino NCE ($\nu$NCE) scattering cross sections~\cite{MB_CCQE, denis_PRD} on carbon.
Recently, a measurement of the neutrino content of the antineutrino mode flux was carried out~\cite{WS_paper} and an antineutrino CCQE ($\bar\nu$CCQE) cross section measurement published~\cite{nubarCCQE}. 
Sizable nuclear effects are observed in both the $\nu$CCQE and $\bar\nu$CCQE data suggesting contributions from nucleon-nucleon correlations and two-body exchange currents~\cite{Amaro_CCQE_2011, Bodek_Budd, Giusti_Meucci_CCQE, Martini_CCQE, Nieves_CCQE, Sobczyk_CCQE}.
The $\nu$NCE cross section has also been studied to quantify these nuclear effects~\cite{Amaro_NCE_superscaling_2006, Benhar_nucelar_effects_NCE_2011, Meucci_NCE_2011, Butkevich_Denis}.
As the RFG model does not incorporate the nuclear effects, agreement between the measured cross sections ($\nu$CCQE, $\nu$NCE, and $\bar\nu$CCQE) and the cross section model was achieved by assigning a higher value ($\sim$30\%) to the axial mass ($M_A$) parameter in the axial-vector from factor. 
For a detailed discussion of neutrino cross section measurements and model predictions see Ref.~\cite{xsec_bible}.

The antineutrino-nucleus NCE ($\bar\nu$NCE) scattering measurement reported here is part of a series of measurements providing an understanding of the neutrino flux and cross sections in the energy regime accessible to MiniBooNE~\cite{MB_CCQE, denis_PRD, ccpi_ratio, ccnpi, CCpi, nubarCCQE}.
The data set corresponds to $10.09 \times 10^{20}$ protons on the neutrino production target.
The experimental signature is the same as the $\nu$NCE scattering~\cite{denis_PRD} -- the scintillation light produced by the recoil nucleons. 
The sample size for the $\bar \nu$NCE scattering cross section -- 60,605 events with 40\% sample purity -- is the largest collected to date for this type of interaction.  
Also reported here are the first experimental measurements of the antineutrino to neutrino NCE cross section and $\bar \nu$NCE to $\bar \nu$CCQE cross section ratios. 

Previously,  a few experiments have measured neutrino-nucleon NCE scattering~\cite{NCE1, NCE2}, most notably the BNL E734 experiment~\cite{bnl734}  which reported both neutrino-proton and antineutrino-proton NCE scattering measurements as a function of four-momentum transfer squared ($Q^{2}$) with 1,686 and 1,821 candidate events respectively.

%
\section{MiniBooNE experiment}\label{sec:mb_experiment}

\subsection{The Experimental setup.}

The MiniBooNE experiment, located at Fermi National Accelerator Laboratory was proposed to test the short base line neutrino oscillations reported by the LSND experiment~\cite{LSND_1997, MB_ocs_improved}. 
In addition, the experiment is well suited to measure a variety of high-statistic, neutrino cross sections~\cite{ccpi_ratio, MB_CCQE, denis_PRD, ccnpi, CCpi}.
It is situated in the Booster Neutrino Beamline (BNB) that produces the neutrino beam via the decay of mesons produced in a proton-beryllium interaction. 
The primary proton beam with a momentum of $8.89\gevc$ is extracted from the Fermilab Booster in 1.6~$\mu$s pulses with $\sim 4\times10^{12}$ protons in each beam pulse.
They impinge on a beryllium target placed in a magnetic focusing horn. 
The p-Be interactions produce a secondary beam of mesons that can be selectively focused or defocused by the magnetic horn. 
In antineutrino mode, the magnetic horn focuses negatively charged particles and defocuses positively charged particles. 
The mesons then decay in an air-filled decay pipe producing a beam of (anti)neutrinos.
Magnetic horn focusing also increases the desired neutrino flux reaching the MiniBooNE detector by a factor of $\sim$ 6.
The average energy of the antineutrino beam is about 650\mev.
Further details on the BNB can be found in Ref.~\cite{MB_flux_paper}. 

The MiniBooNE detector, situated 545\m from the Be target, is a spherical steel tank with a radius of 610\cm filled with 800 tons of mineral oil. 
The mineral oil~\cite{MB_oil} serves both as a target for the neutrino beam and the medium in which the resultant particles from neutrino interactions propagate . 
The detector is divided into two optically isolated regions separated by a spherical shell of radius 575\cm.
The inner sphere, referred to as the \emph{signal} region is lined with 1,280 inward-pointing 8-inch photomultiplier tubes (PMTs)~\cite{MB_PMTs_1}.
The outer shell is the \emph{veto} region with 240 PMTs arranged in back-to-back pairs pointed along the circumference of the detector.
Charged particles produced in the neutrino interaction emit Cherenkov and scintillation light that is collected by the PMTs. 
Six steel legs support the detector, situated in a vault along with the detector electronics and data acquisition (DAQ) systems.
The entire assembly is buried under approximately 3\m of earth overburden to reduce cosmic ray backgrounds.
Further details about the MiniBooNE detector can be found in Ref.~\cite{MB_detector_desc}.

\subsection{The Flux Prediction, Cross Section Model, and Detector Simulation.}\label{sec:flux}
A GEANT4-based Monte Carlo (MC) beam simulation~\cite{GEANT4} is used to calculate the neutrino and antineutrino flux at the detector.
The simulation accepts as input the shape, location and material of the components of the BNB, the MiniBooNE target hall, and the meson decay volume through which the primary protons, the secondary mesons, and tertiary neutrinos propagate.
The various components of simulation depend on the specific processes in the beamline, and arise from a combination of constraints which include: other particle production software, external measurements by MiniBooNE or other experiments in a similar energy regime, theoretical predictions, and extrapolation of external measurements to MiniBooNE energies.

\begin{figure}[t!]
\includegraphics[width=\columnwidth]{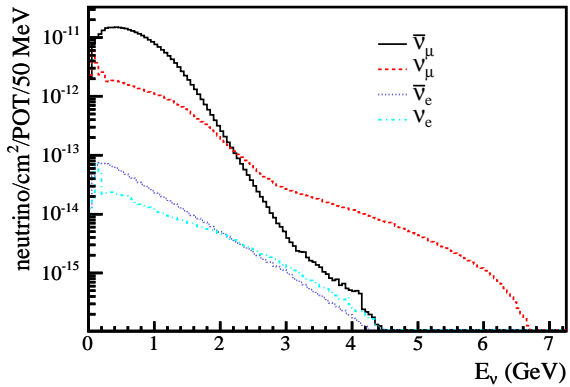}
\vspace{-2mm}
\caption{(color online) The predicted antineutrino mode flux at the MiniBooNE detector for different types of
         neutrinos as a function of their energy as reported in~\cite{MB_flux_paper} and~\cite{flux_table}. }
\label{fig:nubarflux}
\end{figure}

Most of the neutrinos seen by the detector come from the decay of primary $\pi^{+}$ and $\pi^{-}$ produced in the p-Be interaction as well as their subsequent $\mu^{+}$ and $\mu^{-}$ decays.
The $\pi^{+}$ and $\pi^{-}$ production tables used in the MC simulation come from a parametrization of the HARP experiment~\cite{HARP} which measured pion production on a replica Be target at 8.89\gevc. 
The resulting neutrino flux prediction for antineutrino mode running is  shown in Fig.~\ref{fig:nubarflux} (The flux tables are available at~\cite{flux_table}).
The neutrino contamination in the antineutrino mode beam is higher ($\sim$16\%) as compared to the corresponding antineutrino contamination of the neutrino mode beam ($\sim$6\%).
For details on the MiniBooNE flux prediction in both modes see Ref.~\cite{MB_flux_paper}.  
The neutrino contamination in the antineutrino mode beam was measured by the MiniBooNE collaboration~\cite{WS_paper} and the results applied to the flux estimate used in the present measurement. 

Neutrino interaction rates, products, and their kinematics in the MiniBooNE detector are predicted using the {\tt NUANCE}~\cite{Nuance} neutrino event generator that has been customized to the MiniBooNE experiment.
It has as input the neutrino flux prediction described above, as well as the detector target material and geometry.
The mineral oil target is CH$_{2}$ with a density of $0.845\den$.
The (anti)neutrino NCE scattering off of free protons is modeled using the Llewellyn-Smith formalism ~\cite{Llewelllyn-smith}, while for bound nucleons the relativistic Fermi gas (RFG) model of Smith and Moniz~\cite{Smith-Moniz} is used. 
In {\tt NUANCE} pion production is assumed to occur via delta production as per Rein and Sehgal's prescription \cite{ReinSehgal}.

The various parameters in {\tt NUANCE} are tuned as follows:
All the nucleon's vector form factors are assumed to retain their conventional values while the mass in the axial vector $M_{A}$ is assigned a value of 1.23\gev for the nucleons bound in carbon and 1.13\gev for free nucleons.
In order to match the observed MiniBooNE $\nu$CCQE data, particularly in the low $Q^{2}$ regime, the Pauli blocking parameter was scaled up by a scaling factor, $\kappa$=$1.022$ -- see Ref.~\cite{Teppei_thesis} for details.
Note that these values are used for the CCQE/NCE channels that are background to this measurement and the changes as indicated in more recent analyses~\cite{MB_CCQE, denis_PRD} are covered by systematic errors on these parameters.
In the case of neutrino induced resonant pion production, the form factors are assumed to be identical to those used in NCE and CCQE interaction, with the exception of the axial vector mass, where we take $M_{A}$=1.1\gev -- see Ref.~\cite{ccnpi}.
A  20\% probability is assigned to the possibility that the outgoing pion is absorbed within the nucleus through final state interactions (FSI).
In this case the final product of the interaction is just the nucleon, similar to a neutrino NCE interaction.
Lastly, the strange quark contribution to the vector and axial vector form factor is taken to be zero with an uncertainty of 0.1.

The neutrino-generated final states output by the MiniBooNE neutrino event generator ({\tt NUANCE}) are passed on to the MiniBooNE detector simulation. 
A GEANT3 simulation software \cite{GEANT3} in conjunction with a customized optical model is used to simulate particle propagation, the resulting light emission and propagation, and the PMT response in the MiniBooNE detector.
Some modifications to the standard GEANT3 routines include an improved model for Dalitz decay ($\pi^{0} \rightarrow e^{+}e^{-}\gamma$), muon decay ($\mu \rightarrow e\nu\nu$), and  the possibility of $\mu^{-}$ capture by carbon. 
The default GFLUKA~\cite{FLUKA} package is used to model hadron interactions.
The MiniBooNE optical model has twelve components with a total of thirty-five adjustable parameters which have been tuned using external measurements and calibration data. 
The various components of the optical model include: the index of refraction of the oil, light extinction length, the propagation and detection efficiency of Cherenkov light, scintillation and fluorescence yields of the different fluors present in the oil, scattering and reflections in the detector, and relative and angular efficiencies of the PMTs.
The scintillation photons were modelled as per Birk's law~\cite{birks_law} and its coefficients are additional parameters in the optical model. 
The charge and time response of the PMTs were modelled by parametrization of data collected by the PMT studies using a pulsed laser source and calibration light sources in the detector~\cite{MB_PMTs_1}. 
Finally, the detector simulation includes modelling of signal digitization of the PMT outputs and the data acquisition.

%
\section{Neutral--Current Elastic Analysis.}\label{sec:nce_analysis}
\subsection{Event Reconstruction.} 

\begin{figure}
\includegraphics[bb = 0 0 568 385, width=\columnwidth]{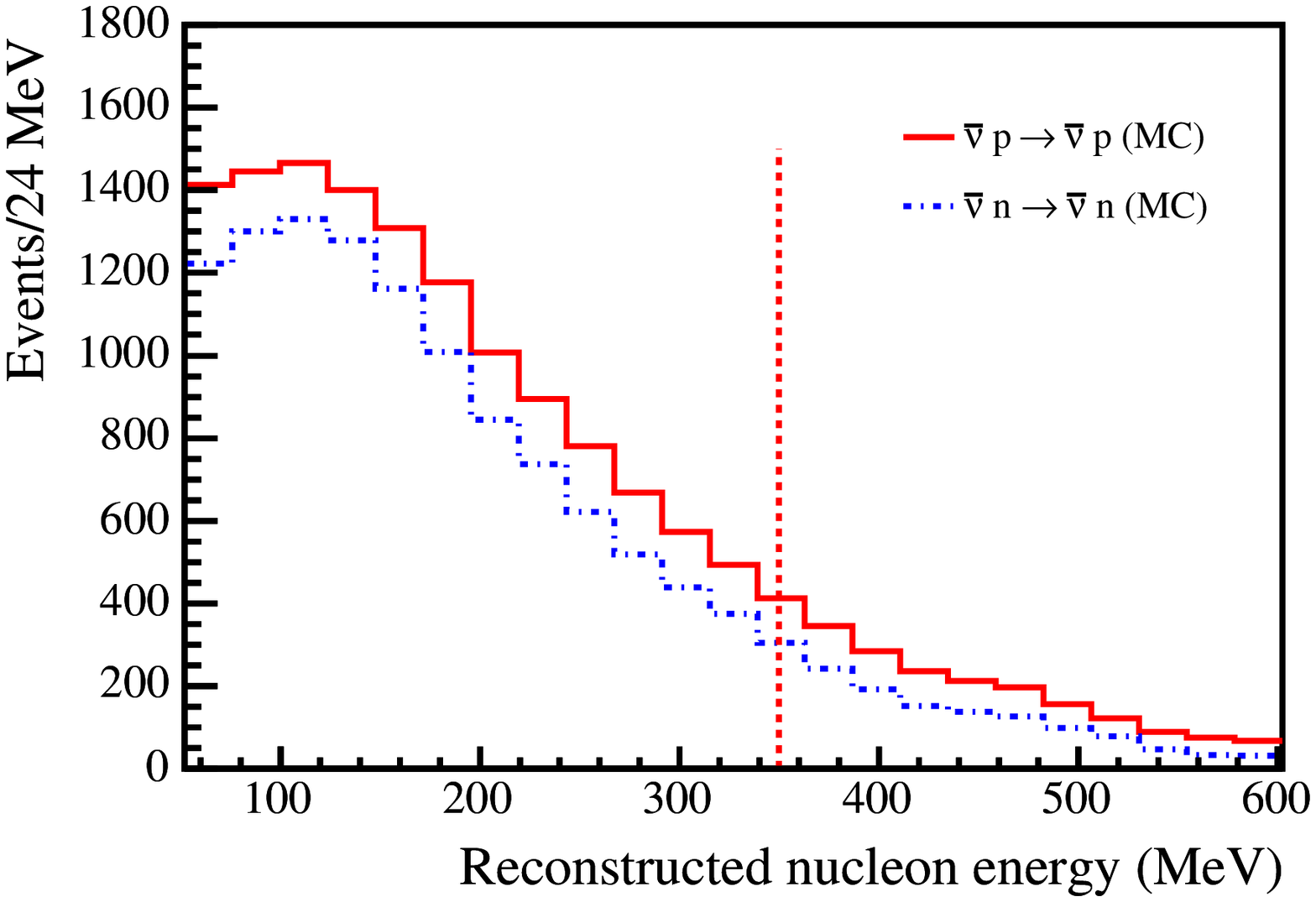}
\includegraphics[bb = 0 0 568 385, width=\columnwidth]{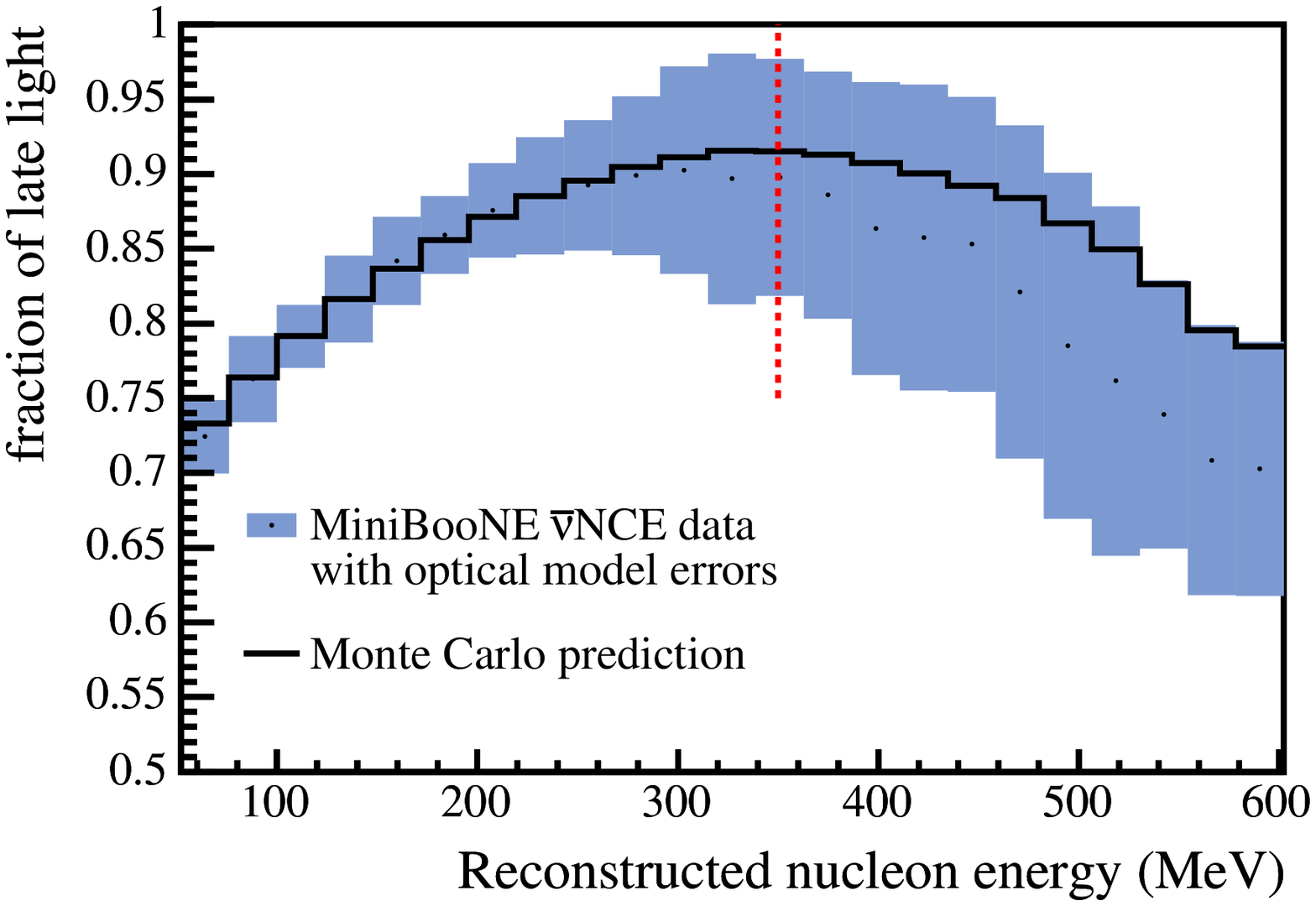}
\caption{(color online) $\bar \nu$NCE reconstruction in MiniBooNE. The top plot shows the MC predicted energy spectrum for $\bar \nu$NCE protons and neutrons. We see that most of the scattered nucleons are below the Cherenkov threshold for protons in the MiniBooNE medium (350\mev -- shown by the dotted red line) and that both neutrons and protons have a similar spectrum. The bottom plot shows the fraction of late (scintillation) light for $\bar \nu$NCE data and MC, as a function of reconstructed nucleon energy. The errors on data represent the uncertainty in the modelling of the optical photons resulting from the scintillation light.  We see that there is good agreement between the prediction and data from 50\mev to 350\mev and also the Cherenkov threshold transition matches at 350\mev (shown by the dotted red line). }
\label{fig:scifra}
\end{figure}

In the case of NCE event reconstruction, each event is assumed to be due to a proton whose Cherenkov and scintillation light profiles are determined from the MC simulation. 
NCE scattering resulting in outgoing neutrons is only seen through their subsequent strong interactions resulting in protons, hence NCE neutrons are indistinguishable from NCE protons.
Most NCE protons are below Cherenkov threshold  (350\mev) and are reconstructed primarily via the scintillation light yields. 
Figure~\ref{fig:scifra}(top) shows the MC prediction of reconstructed  energy spectrum for NCE protons and neutrons.
We see that most of the scattered nucleons are below threshold and that NCE neutrons and NCE protons have a similar energy profile.

\begin{figure}
\includegraphics[bb = 0 0 568 385, width=\columnwidth]{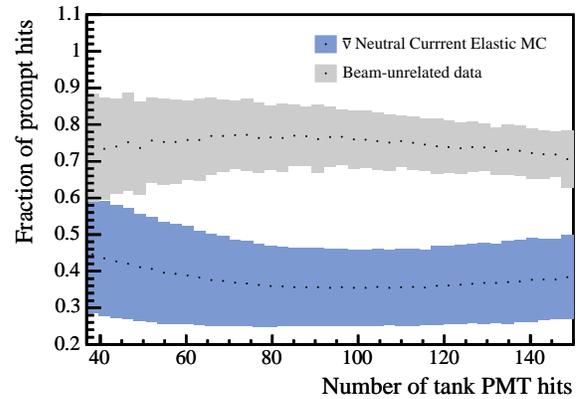}
\caption{(color online) Fraction of prompt photons compared to the total number for
         beam unrelated events and NCE MC events reconstructed under an electron
         hypothesis. The events out of time with the beam are due to Michel decay while the events in time with the beam are due to scintillation light from recoiling nucleons. 
         The error bars correspond to the RMS of the distributions.}
\label{fig:elfra}
\end{figure}

The charge and time information from the PMTs is used to determine the position, time, direction and energy of an event by employing a log-likelihood minimization method. 
Outgoing protons from NCE scattering have a characteristic light emission profile which readily allows for their particle identification. 
The ability to differentiate protons from beam unrelated events (mostly electrons)  is illustrated in Fig.~\ref{fig:elfra} where the fraction of prompt light emitted is plotted versus the number of tank PMT hits.
Prompt light is defined as the fraction of PMT hits with corrected time between -5 and 5 ns, where the corrected time is the time difference between the PMT hit time and the reconstructed event time with light propagation time from the reconstructed vertex to the PMT taken into account.

The MiniBooNE detector position resolution for NCE protons is $\sim$ 0.75\m and $\sim$1.35\m for neutrons.
The energy resolution is $\sim$20\% for protons and $\sim$30\% for neutrons. 
Energy scaling for NCE protons was checked by plotting the fraction of scintillation or late light as a function of reconstructed energy, as shown in Fig.~\ref{fig:scifra} (bottom).
 There is agreement (within errors) between data and MC both in the energy regime of interest (50\mev to 350\mev) and the Cherenkov threshold transition at 350\mev.
For details on reconstruction methods used in MiniBooNE, see Ref.~\cite{MB_reconst}; for the NCE event reconstruction in particular see Ref.~\cite{denis-thesis}. The energy calibration of NCE protons is discussed in Appendix C of Ref.~\cite{denis-thesis}.

%
\subsection{Event Selection.}\label{sec:cuts}
\begin{figure}
\includegraphics[scale=0.3]{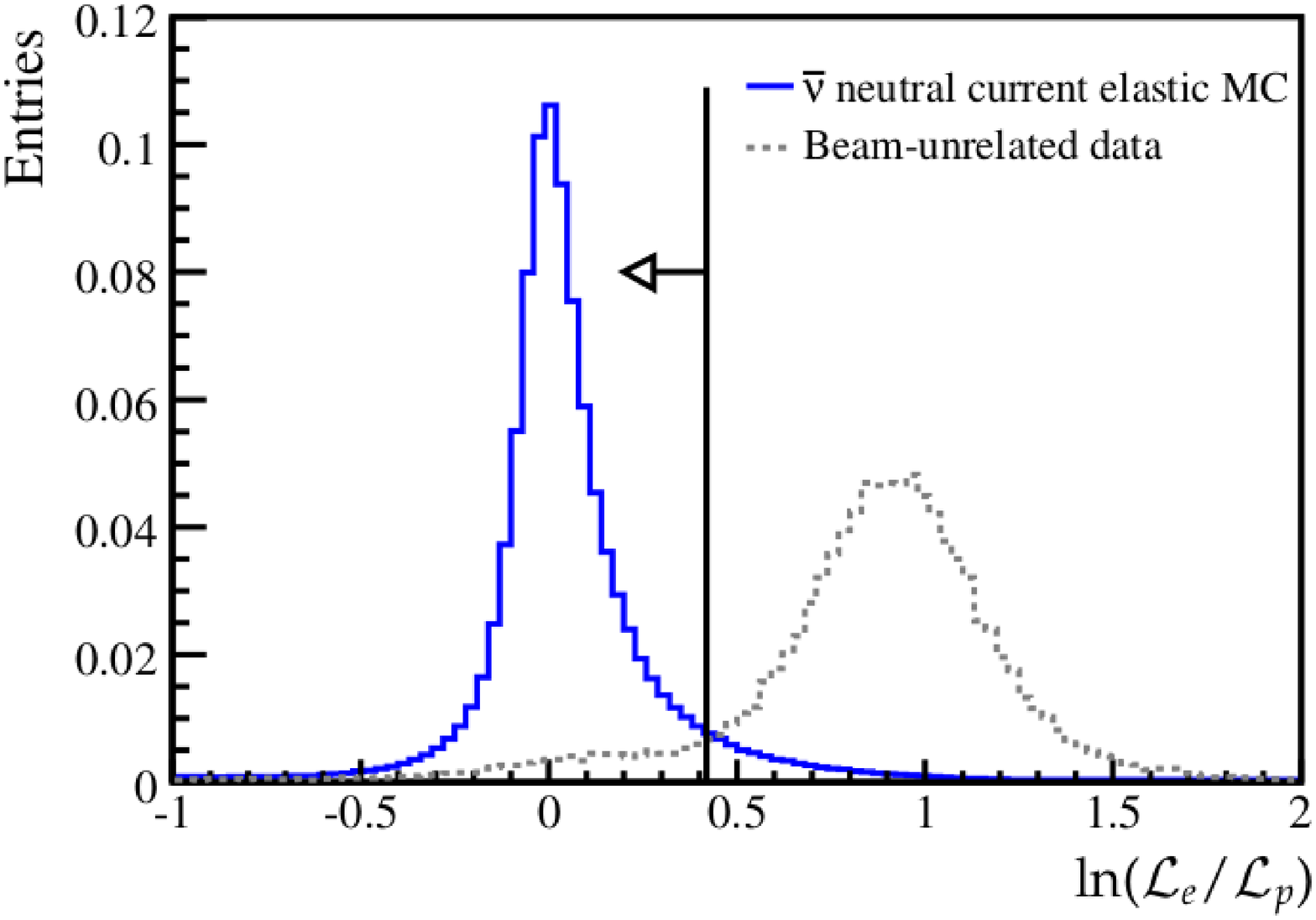}

\caption{(color online) Log-likelihood ratio between electron and proton event hypotheses for
         MC-generated NCE scattering events and beam unrelated data.
         Both histograms are normalized to unit area.
         Events with $\ln(\mathcal{L}_e/\mathcal{L}_p)<0.42$ are selected for the analysis.}
\label{fig:tllkdiff}
\end{figure}

In order to isolate a sample of $\bar\nu$NCE events a series of analysis cuts based on  the physics processes and Monte Carlo studies was applied. 
The cuts (listed below) are various restrictions on the experimental variables, like PMT charge, time or reconstructed energy, which differentiate the NCE events from other events. 
\begin{enumerate}
\item Only 1 \emph{subevent} to ensure selection of NC events with no decaying particles.
A subevent is a cluster of at least 10 tank hits with no more than 10 ns between any two consecutive hits. 
A typical NCE interaction has only one subevent associated with the primary neutrino interaction.

\item Number of veto PMT hits less than 6.
This cut excludes events that are entering or exiting the detector and register activity in the veto region.
Cosmic rays and neutrino interactions in the material surrounding the detector with the outgoing nucleon entering the detector account for most of the events constrained by this selection cut.
The veto cut removes almost all (99.9\%) of the cosmic ray background. 
CCQE interactions in which the muon exits the detector before decaying are also excluded by this cut.

\item The reconstructed event time must occur within the neutrino beam time window.

\item Number of  PMT tank hits greater than 12 to ensure that the event can be reliably reconstructed. 

\item Reconstructed proton energy less than 650\mev, above which the signal to background ratio decreases significantly. 

\item A cut on the log-likelihood ratio between events reconstructed with a proton hypothesis and an electron hypothesis:\ $\ln(\mathcal{L}_{e}/\mathcal{L}_{p})<0.42$.
This cut removes beam-unrelated (Michel) electrons from cosmic ray muon decays.
Fig.~\ref{fig:tllkdiff} shows the likelihood difference between events reconstructed under an electron and a proton hypothesis, for both Monte Carlo $\bar \nu$NCE scattering events and beam-unrelated backgrounds (data).

\item Finally, a fiducial volume cut of 5\m.
This cut ensures that the events in the sample are well-reconstructed and well-contained.
It also reduces the neutrino events resulting from interaction with the earth surrounding the detector. 

\end{enumerate}

\begin{figure}
\includegraphics[bb = 0 0 568 385, width=\columnwidth]{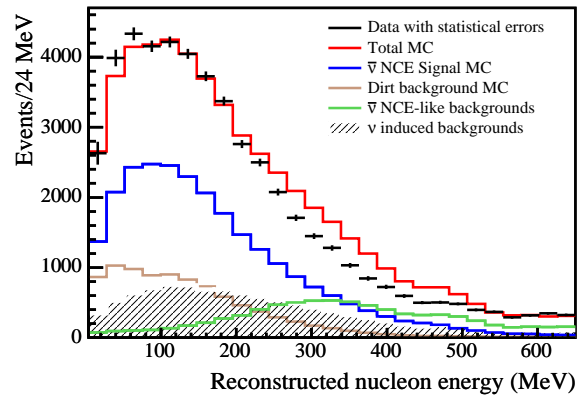}
\caption{(color online) Reconstructed nucleon kinetic energy spectra for the data and MC
         after the NCE event selection and a uniform fiducial volume cut of $R<5$~m are applied.
         All MC distributions are normalized to the number of protons on target (POT).}
\label{fig:erec}
\end{figure}

A total of 60,605 events pass the analysis cuts, representing the largest $\bar\nu$NCE candidate sample ever collected to date. 
Table ~\ref{tab:eventsample} shows the results of a MC study to determine the efficiency and purity of $\bar \nu$NCE sample for each selection cut applied in the order shown.
Figure~\ref{fig:erec} shows the reconstructed nucleon energy spectrum for NCE events (data) along with the MC prediction of the sample composition, after subtracting beam-unrelated events and estimation of backgrounds (next section).
After removing beam-unrelated events, the predicted fraction of $\bar\nu$NCE scattering events in the sample is 48\%. 
The remaining 52\% of events are various backgrounds to this measurement.
Neutrino induced interactions constitute 19\% of the background.
The next largest source of background are the so called ``dirt events''(17\%). These are neutrino interactions happening in the earth just outside the detector with the recoil nucleon entering the detector without firing enough veto PMTs.
Finally, there is a contribution from NCE-like events (14\%), which are NC-pion producing events where the pion is absorbed in the target nucleus resulting in an event with a nucleon mimicking the neutrino NCE scattering signal.
 
\begin{table}
\begin{ruledtabular}
\begin{tabular}{lcc}

Selection cut &  efficiency &  purity \\
\hline\noalign{\smallskip}

No cuts                                                            & 100\%         & 0.2\% \\
1 subevent \&  Veto hits $<6$                        & 59\%            & 2\% \\
Event in beam window                                     & 57\%           & 15\% \\
Tank hits $>12$                                             & 55\%           &  16\% \\
 Energy $<650$\mev                                      & 45\%           &  16\% \\
$\ln(\mathcal{L}_{e}/\mathcal{L}_{p})<0.42$   & 42\%           & 34\%  \\
Fiducial cut $R<5$\m                                      & 32\%           & 40\% \\

\end{tabular}
\end{ruledtabular}
\caption{\label{tab:eventsample} Results from a MC study for $\bar\nu$ NCE efficiency and purity as a function of the selection cuts.}

\end{table}

%
\subsection{Estimation of Backgrounds}\label{sec:Background estimation}
The various backgrounds to the MiniBooNE $\bar\nu$ mode NCE measurement are similar to those in the $\nu$ mode NCE measurement~\cite{denis_PRD} with one notable exception -- the neutrino induced events in the antineutrino mode beam.  
Data-driven methods are used to constrain all of the backgrounds as explained below.

As previously mentioned, the neutrino contamination in the antineutrino flux ($\sim$16\%) is significantly larger than the corresponding antineutrino contamination in neutrino mode ($\sim$4\%).
HARP~\cite{HARP} did not cover all the phase space for $\pi^{+}$ production necessary to specify the neutrino background in the antineutrino beam.
The neutrino contamination in the antineutrino mode beam was measured by MiniBooNE to constrain the flux outside of the region where data from the HARP experiment are available.
Three independent and complimentary techniques are employed to measure the neutrino background utilizing the high-statistic neutrino mode cross section measurements made in an almost pure neutrino beam. 
Briefly, the first technique exploits the difference in the angular distribution of outgoing muons in $\bar\nu$ and $\nu$ CCQE interactions to tag neutrino induced events.
The second technique inferred the rate of $\nu$ induced events from a study of the charged current (CC) single pion production channel.
 The neutrino CC single pion interaction leads to a $\pi^{+}$ whose decay into a muon is seen in the detector, however the corresponding antineutrino CC single pion interaction produces a $\pi^{-}$ that is absorbed in the detector medium most of the time~\cite{pion_capture}.
The third technique to constrain the $\nu$ component of the beam exploits an external measurement of the rate of $\mu^{-}$ nuclear capture in $\nu$ CC interactions~\cite{muoncap}.
The results from the three techniques are consistent.
For details on the first two techniques see Ref.~\cite{WS_paper} and for the third technique see Appendix A in Ref.~\cite{nubarCCQE}.
Accordingly, a correction factor of $0.78$ was applied to the MC prediction of the original neutrino flux  (in the antineutrino mode) based on the HARP measurement.
The antineutrino flux prediction was unchanged in this procedure.
The cross section of neutrino induced NCE events in the antineutrino mode is inferred from the high-statistic neutrino mode $\nu$NCE cross section measurement~\cite{denis_PRD}.
The total uncertainty in the estimation of the $\nu$NCE background events in the sample is 14\%.

\begin{figure}[t!]
\includegraphics[width=\columnwidth]{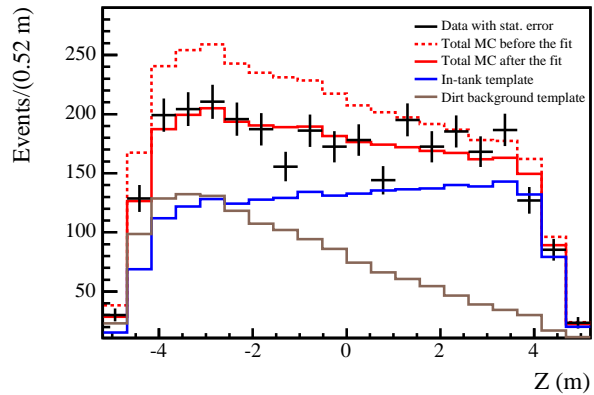}
\vspace{-2mm}
\caption{(color online) Fit to the data using MC templates for the in-tank and dirt events in reconstructed Z variable for energy between 40\mev and 110\mev. The ``total MC before the fit'' is a sum of the in-tank and dirt templates, which are absolutely (POT) normalized.}
\label{Z_dirt}
\end{figure}

The next major background are neutrino/antineutrino interactions occurring in the earth surrounding the detector resulting in nucleons (mostly neutrons) which penetrate the detector without firing veto PMTs. 
The so called dirt events are a significant fraction of the total background, particularly at low (below 300 MeV) energies.
They are difficult to model as they result from interactions with various media outside the detector (the soil, detector support structures etc.), whose exact composition is not known.
However, they have distinct kinematics and spatial distributions that can be used to constrain their contribution.
Dirt events are mostly low in energy, preferentially reconstructed in the upstream part (the side facing the beam), and close to the edge of the detector.
We use these kinematic information to select a sample of ``dirt-enriched'' events in both data and MC.
A chi-square minimization is then employed to fit the MC prediction with a single scale factor to the observed data.
The resulting fits show agreement across the three variables which were chosen: the reconstructed Z variable (axis along the beam direction), reconstructed R variable (radius) and reconstructed energy.
Figure~\ref{Z_dirt}  shows a representative energy bin in a sample enriched in ``dirt events'' produced with the reconstructed Z variable. 
For details on the dirt measurement method used in MiniBooNE NCE analyses see Appendix A in Ref.~\cite{denis_PRD}. 
This background was separately measured in this $\bar \nu$NCE analysis and $\nu$NCE analysis~\cite{denis_PRD}, as the dirt background composition could be different in the two cases.
The resultant dirt scaling factor of 0.62 was applied to the MC dirt prediction with an uncertainty of 10\%. 
This may be compared to the scaling factor of 0.68 for the neutrino mode~\cite{denis_PRD}.

The final background is the $\bar\nu$NCE-like events which are NC-pion events where the pion is absorbed within the nucleus, resulting in a final state identical to a $\bar\nu$NCE scattering signal event.
We rely on the MC prediction (together with the MiniBooNE measured neutral current $\pi^{0}$ cross section~\cite{colin_paper}) to estimate this background. 
The {\tt NUANCE} cross section model assigns an error of 30\%  for pion absorption. 
Note that the $\bar\nu$NCE-like background is the only background not directly measured in this analysis, hence we report the calculated contribution which was subtracted to obtain our final antineutrino-nucleon NCE scattering cross section measurement (Fig.~\ref{fig:xs}).  
Table~\ref{tab:backgrounds} lists the MiniBooNE $\bar\nu$NCE analysis sample composition after estimation of backgrounds.

\begin{table}
\begin{ruledtabular}
\begin{tabular}{lc}

Sample composition &  Fraction  \\
\hline\noalign{\smallskip}

$\bar\nu$NCE (signal)          & 49\%          \\
$\nu$NCE (background)        &19\%           \\
Dirt events                            & 17\%            \\
$\bar\nu$NCE-like               & 13\%            \\
Others                                  &   2\%        

\end{tabular}
\end{ruledtabular}
\caption{\label{tab:backgrounds} The MiniBooNE $\bar\nu$NCE analysis sample composition after backgrounds estimation.}
\end{table}

\subsection{Antineutrino Neutral--Current Elastic Flux-Averaged Cross Section.}\label{sec:xs}

 We report the $\bar\nu$NCE scattering differential cross section as a function of quasi-elastic momentum transfer $(Q_{QE}^{2})$. 
For both CCQE and NCE scattering, $(Q_{QE}^{2})$ is determined assuming a quasi-elastic scattering of a neutrino off an at-rest nucleon.
However, to calculate $(Q_{QE}^{2})$  the outgoing muon kinematics is used in case of CCQE scattering  whereas, in case of NCE scattering we use the recoil nucleon.  
According to our final-state interaction model the total kinetic energy of all of the outgoing nucleons in an NCE interaction is a good proxy for the primary outgoing nucleon and to   $Q_{QE}^{2}$.
And the sum of the kinetic energies of all final state nucleons produced in an NCE interaction is proportional to the total charge on all PMTs. 
\begin{equation*}
Q^{2}_{QE}=2m_{N}T=2m_{N}\sum_{i}T_{i},
\end{equation*}
where $m_{N}$ is the nucleon mass and $T$ is the sum of the kinetic energies of the final state nucleons.
The sum $T_{i}$ is used in the definition due to the calorimetric nature of the MiniBooNE measurement and is more inclusive with respect to possible nuclear effects as compared to track-based reconstruction used in the SciBooNE experiment~\cite{sciboone_track} or the BNL E734 experiment~\cite{bnl734}.
 
The constrained backgrounds--the beam unrelated background, the dirt background, and the neutrino induced backgrounds--are subtracted from the reconstructed energy spectrum for data.
The $\bar\nu$NCE-like background is removed by doing a bin-by-bin multiplication of the data spectrum by the signal fraction, i.e. the ratio of number of $\bar\nu$NCE events to the total number of antineutrino induced in-tank events, based on the MC prediction.
Finally, a Bayesian unfolding procedure~\cite{cowan_book, bayes_unf} is used to correct the background subtracted data for limited detector resolution, mis-reconstruction, and sources of detector inefficiency.

The flux-averaged $\bar\nu$NCE scattering differential cross section is extracted as per the formula:
\begin{equation} 
\frac{d\sigma_{i}^{\bar\nu NCE}}{dQ_{QE}^{2}}=\frac{\sum_{j}U_{ij}(d_{j}-D_{j}-V_{j}-N_{j})\frac{S_{j}}{S_{j}+B_{j}}}{ \epsilon_{i}\cdot (2 M_{N} \Delta T) \cdot N^{tar}\cdot N^{POT}\cdot \Phi_{\bar\nu}}
\end{equation} 
where $U_{ij}$ is the unfolding matrix, the index $j$ labels the reconstructed energy bin, and $i$ labels the unfolded true energy bin (as per the MC prediction).
In the above equation, $d_{j}$ represents data, $D_{j}$ and $N_{j}$ are the data-driven corrected backgrounds of dirt events and beam unrelated events respectively, $S_{j}$ is the MC predicted number of $\bar\nu$NCE scattering events, $B_{j}$ is the rest of the backgrounds which mostly consists of the $\bar\nu$NCE-like events, $\epsilon$ is the efficiency, $\Delta T$ is the bin width, $N^{tar}$ is the number of nucleons in the detector, $N^{POT}$ is the number of protons on target corresponding to the data set, and $\Phi_{\bar\nu}$ is the total integrated antineutrino flux within the energy range 0 to 10 GeV (both $\bar\nu_{\mu}$ and $\bar\nu_{e}$).

The unfolding matrix $U_{ij}$ is calculated from the predicted correlation between reconstructed nucleon energy and true nucleon kinetic energy (the sum of the kinetic energies of all nucleons in the final state) resulting in a well-behaved, but biased solution.
The error due to the bias in the unfolding procedure is estimated by employing an iterative method where each successive unfolded spectrum is used as the true energy spectrum for the next iteration. 
The resulting spread in the cross section measurement, from the first iteration to the last one when it converges, is the error in the unfolding procedure. 
The details of the unfolding procedure and the estimation of the associated error with it can be found in Refs.~\cite{denis-thesis, mythesis}.
\begin{figure}
\includegraphics[scale=0.46]{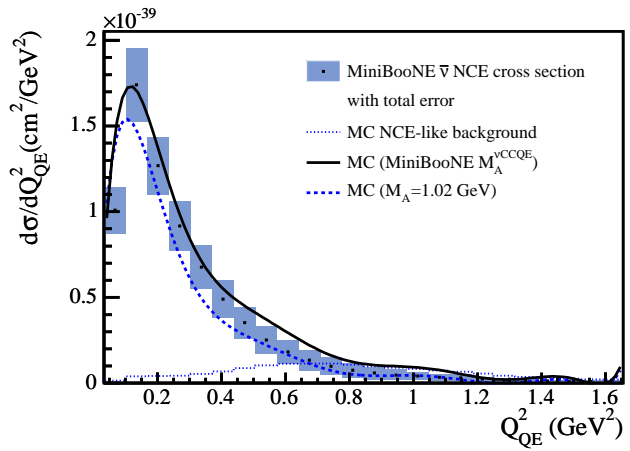}
\caption{(color online) MiniBooNE $\bar \nu N \rightarrow \bar \nu N$ flux-integrated differential cross section on CH$_2$. The uncertainty includes all errors -- systematic and statistical. Also shown is the NC-like background which has been subtracted from the reported cross section. The Monte Carlo predictions for the cross section with two different values of $M_{A}$ -- MiniBooNE $\nu$CCQE measured value, $M_A=1.35$\gev  (black) and $M_A=1.02$ \gev (dashed blue) --  are shown for comparison.}
\label{fig:xs}
\end{figure}
The resulting flux-averaged $\bar\nu$NCE scattering differential cross section is shown in Fig.~\ref{fig:xs}.
Also shown is the $\bar\nu$NCE-like background which was subtracted from the total $\bar\nu$NCE-like cross section.  
Though the systematic uncertainties are higher in the lowest energy bins, as $Q^{2}$ approaches $0$, the ``roll-over'' associated with the binding energy of the carbon nucleus is clearly seen for the first time.

Since the MiniBooNE target is mineral oil (CH$_{2}$), the measured $\bar \nu$NCE scattering is a sum of three different processes: scattering on free protons in hydrogen, bound protons in carbon, and bound neutrons in carbon.
The contribution of these individual processes to the total cross section is discussed in Appendix A. 
Integrating over $0.033\gevsq$ to $1.655\gevsq$ $Q^{2}_{QE}$ bins, the total $\bar\nu$NCE scattering cross section per nucleon is $(5.06 \pm 0.990) \times 10^{-40}\cmsq$.

The various uncertainties in  the $\bar \nu$NCE cross section measurement are listed in Table~\ref{tab:error}. 
The flux error encompasses the uncertainties in the pion propagation and decay in the BNB.
The cross section error includes the uncertainties in the cross section model of the various background processes.
The error associated with the detector electronics, the PMT response, and the uncertainty in modeling the production and propagation  of optical photons within the detector medium contribute to the detector error.
The next two errors are from the uncertainty in the measurement of neutrino induced NCE events and ``dirt'' events.
The final error is due to the unfolding procedure.

For each systematic uncertainty, there is an associated error matrix that encompasses the information about the parameters describing the particular physical process, the uncertainties in the parameters, and any correlation among them.
The error matrices are added in quadrature to obtain the total error matrix and Table~\ref{tab:error} lists the normalization error (sum of  the diagonal elements of the error matrix) for both the individual  errors and the total error. 

\begin{table}
\begin{ruledtabular}
\begin{tabular}{lc}
\quad{Error}                                                            & Value (\%) \\ 
\hline\noalign{\smallskip}
Statistical                                                              &4.5 \tabularnewline
Flux uncertainty                                                     &5.8\tabularnewline
Cross section uncertainty (background processes) & 3.0 \tabularnewline
Detector effects                                                     & 14.5 \tabularnewline
Estimation of $\nu$ induced events                       &3.6 \tabularnewline
Estimation of ``dirt'' events                                  & 1.7 \tabularnewline
Unfolding error                                                      &6.8 \tabularnewline

\hline\noalign{\smallskip}
\textbf{\quad Total}			   		 & 19.5\tabularnewline
\end{tabular}
\end{ruledtabular}
\caption{\label{tab:error}The total integrated normalization error in the MiniBooNE $\bar \nu$NCE scattering cross section measurement along with the key individual error contributions. 
}
\end{table}

\subsection{Antineutrino Neutral--Current Elastic to Neutrino Neutral--Current Elastic Cross-Section Ratio Measurement.}\label{sec:nce_ccqe_ratio}
Both the neutrino-nucleus NCE scattering cross section~\cite{denis_PRD} and the antineutrino-nucleus NCE cross section reported here represent the largest sample of such events ever collected to date.
Since both measurements were made in the same beamline and with the same detector, we expect a bin-by-bin ratio of the two cross section measurements would cancel the common systematic errors. 
The resulting cross section ratio plot encompasses information from both the neutrino and antineutrino NCE scattering cross sections while minimizing the errors.
However, it should be noted that $Q^{2}_{QE}$ is sensitive to the neutrino flux and the two measurements are made in the same beamline but with opposite horn polarities, resulting in non-identical flux spectra.
One of the main motivations for measurement of this cross section is to better understand and model neutrino nucleus interactions. We believe that such a ratio measurement where the errors are carefully accounted for would aid the theoretical physics community to test various models. 

The data set for the ratio measurement consists of the entire neutrino mode and antineutrino mode NCE scattering cross section data from MiniBooNE.  This consists of 94,531 $\nu$NCE candidate events  and 60,605 $\bar\nu$NCE candidate events that pass selection cuts.

\begin{figure}[t!]
\includegraphics[width=\columnwidth]{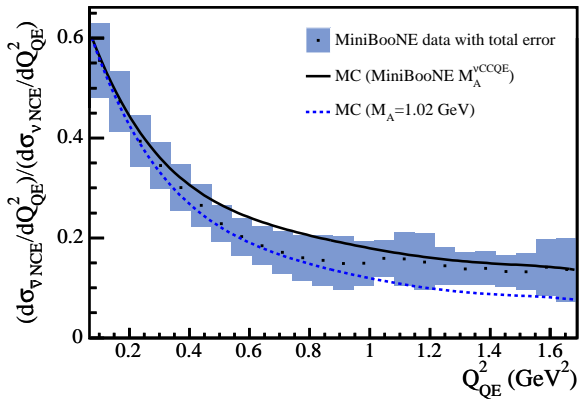}
\vspace{-2mm}
\caption{(color online) Ratio of the antineutrino to neutrino NCE scattering cross section in MiniBooNE with total error. Also plotted are the predicted ratios from MC simulations with $M_A=1.02$\gev, and $M_A=1.35$\gev -- determined by the MiniBooNE $\nu$CCQE measurement ($M_{A}^{\nu CCQE}$) .
}
\label{fig:nce_ratio}
\end{figure}

The systematic error for the ratio measurement was evaluated by dividing the errors into two types: \emph{correlated} errors and \emph{uncorrelated} errors. 
The correlated systematic errors are common to both $\nu$NCE and $\bar \nu$NCE scattering measurements.
Since both measurements are made using the same detector and have the same observed final state, the detector systematic errors -- the uncertainty in the optical photon production and propagation, the error associated with the detector electronics, and the error associated with the PMT response -- are  categorized as correlated errors.  
The uncorrelated errors include the error associated with the measurement of the dirt background, the error in the measurement of the neutrino component of the antineutrino beam, and the error accrued due to the bias in the unfolding procedure implemented. 
The resulting $\bar \nu$NCE to $\nu$NCE cross section ratio measurement is shown in Fig~\ref{fig:nce_ratio}. 
The error bars represent the total normalization error due to both systematic and statistical errors.
The total uncertainty in the $\bar\nu$NCE  to $\nu$NCE cross section ratio measurement is about 20\%.

\subsection{Antineutrino Neutral--Current Elastic to Antineutrino Charged-Current Quasielastic Cross-Section Ratio Measurement.}\label{sec:nce_ccqe_ratio}
\begin{figure}[t!]
\includegraphics[width=\columnwidth]{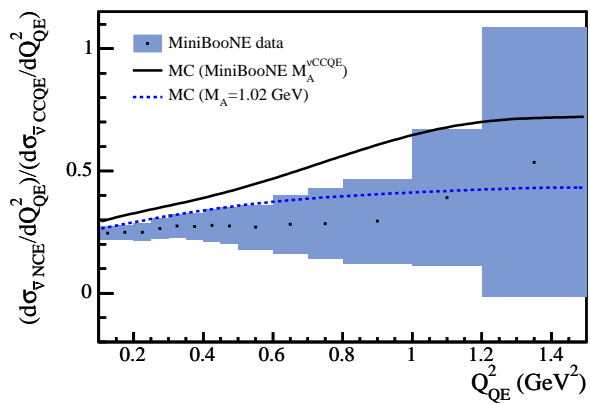}
\vspace{-2mm}
\caption{(color online) MiniBooNE $\bar \nu$NCE/$\bar \nu$CCQE cross section ratio on CH$_{2}$ as a function of $Q^{2}_{QE}$. Also shown are the MC prediction for the ratio with $M_A=1.02$\gev,  as well as that determined by the MiniBooNE $\nu$CCQE measurement ($M_{A}^{\nu CCQE}$=1.35\gev). The individual $\bar \nu$NCE and $\bar \nu$CCQE cross sections are per target nucleon -- there are 14/8 times more target nucleons in the numerator than in the denominator. The error bars include both statistical and systematic errors (except the flux errors) taken in quadrature.}
\label{fig:ccqe_ratio}
\end{figure}

We also extract a $\bar \nu$NCE to $\bar \nu$CCQE scattering ratio measurement in terms of $Q_{QE}^{2}$. MiniBooNE has previously reported this ratio in neutrino mode ($\nu$NCE to $ \nu$CCQE ratio) in Ref.~\cite{denis_PRD}.
It should be noted that there are significant differences between the extraction of  the $\bar \nu$NCE and the $\bar \nu$CCQE scattering cross sections.
In both cases, we assume a stationary nucleon, but in the case of $\bar \nu$CCQE, the momentum transfer $Q^{2}_{QE}$ is determined from the kinematics of the outgoing muon ($\mu^{+}$), whereas in the case of $\bar \nu$NCE $Q^{2}_{QE}$ is calculated from the sum of the kinetic energies of the final state nucleons.
In the $\bar \nu$NCE/$\bar \nu$CCQE scattering cross section ratio (Fig.~\ref{fig:ccqe_ratio}), the uncertainties in the flux estimation are assumed to cancel out whereas other errors have been added in quadrature.

\section{Summary.}\label{sec:Summary}
To summarize, using a high-statistics sample of $\bar \nu$NCE scattering interactions collected by the MiniBooNE experiment the $\bar\nu$NCE ($\bar\nu N \rightarrow \bar\nu N$) flux-averaged differential cross section, $d\sigma/dQ^{2}$ on CH$_{2}$ was measured.
For the first time, an antineutrino to neutrino NCE scattering cross section ratio has also been reported, that accounts for all the systematic errors common to both measurements. 
Finally, the $\bar \nu$NCE to $\bar \nu$CCQE cross section ratio is provided. The corresponding neutrino mode ratio  was reported in Ref.~\cite{denis_PRD}, facilitating a comparison between the two modes.
Keeping in mind the different neutrino flux of the individual cross sections, these cross section ratio measurements are arguably independent of various nuclear effects common to the individual cross sections, providing additional tools for model testing.

The $\bar\nu$NCE cross section (Fig.~\ref{fig:xs})
shows good agreement to a simple RFG model with $M_A=1.35$\gev -- determined by the MiniBooNE $\nu$CCQE measurement~\cite{MB_CCQE}.
This is interesting as it shows that a simple tuning of $M_A$, presumably to effectively handle more complex nuclear effects, provides a reasonable description.
Any other models designed to explain the MiniBooNE $\nu$CCQE data~\cite{Amaro_CCQE_2011, Bodek_Budd, Giusti_Meucci_CCQE, Martini_CCQE, Nieves_CCQE, Sobczyk_CCQE} need to consider the entire MiniBooNE data set ($\nu$CCQE, $\bar\nu$CCQE, $\nu$NCE, and $\bar\nu$NCE) to be considered complete.
 
The authors would like to acknowledge the support of Fermilab, the
Department of Energy, and the National Science Foundation in the
construction, operation, and data analysis of the Mini Booster Neutrino
Experiment.  

\appendix
\section{MiniBooNE Antineutrino Neutral--Current Elastic Cross-Section Discussion.}
\begin{figure}[t!]
\includegraphics[width=\columnwidth]{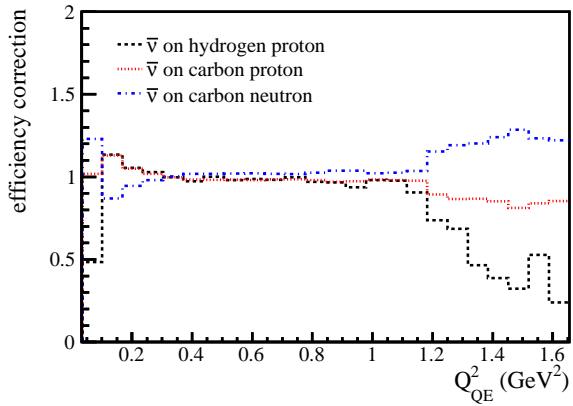}
\vspace{-2mm}
\caption{(color online) The efficiency corrections for the different processes (as labeled) contributing to the MiniBooNE $\bar \nu$NCE scattering cross section,  as a function of $Q^{2}_{QE}$.}
\label{effcorr}
\end{figure}

The antineutrino-nucleon NCE scattering cross section reported here is in terms of $Q^{2}$ which  in MiniBooNE, is proportional  to the total kinetic energies of all the final state nucleons that are produced in the interaction. 
Also in MiniBooNE, NCE scattering  on protons are indistinguishable from NCE scattering on neutrons as the neutrons are seen only via their subsequent strong interaction with protons. 
The MiniBooNE target is mineral oil (CH$_{2}$), hence the  scattering is off of both bound nucleons (in carbon) and free nucleons (in hydrogen).
In fact, the cross section is a sum of  three different processes: the antineutrino scattering off free protons in the hydrogen atom, the bound protons in the carbon atom and, the bound neutrons in the carbon atom. 
Each of the individual processes have  different efficiencies in the MiniBooNE detector. Fig.~\ref{effcorr} shows the efficiency correction functions $C_{\bar \nu p, H}$, $C_{\bar \nu p, C}$, and $C_{\bar \nu n, C}$ for the three processes. The efficiency correction is defined as the ratio of the efficiency for a particular type of  $\bar \nu$NCE event to the average efficiency for all $\bar \nu$NCE events as a function of $Q^{E}_{QE}$. Therefore, flux-averaged $\bar \nu$NCE differential cross section on \ch2 shown in Fig.~\ref{fig:xs} can be expressed as:
\begin{eqnarray*}
\frac{d\sigma_{\bar \nu N \rightarrow \bar \nu N}}{dQ^{2}}=\frac{1}{7}C_{\bar \nu p, H}(Q^{2}_{QE})\frac{d\sigma_{\bar \nu p \rightarrow \bar \nu p, H}}{dQ^{2}} \\
+ \frac{3}{7}C_{\bar \nu p, C}(Q^{2}_{QE})\frac{d\sigma_{\bar \nu p \rightarrow \bar \nu p, C}}{dQ^{2}}\\
+\frac{3}{7}C_{\bar \nu n, C}(Q^{2}_{QE})\frac{d\sigma_{\bar \nu n \rightarrow \bar \nu n, C}}{dQ^{2}}
\end{eqnarray*}
where $d\sigma_{\bar \nu p \rightarrow \bar \nu p, H}/dQ^{2}$ is the $\bar \nu$NCE cross section on free protons (per free proton), $d\sigma_{\bar \nu p \rightarrow \bar \nu p, C}/dQ^{2}$ is the $\bar \nu$NCE cross section on bound protons (per bound proton), and $d\sigma_{\bar \nu n \rightarrow \bar \nu n, C}/dQ^{2}$ is the $\bar \nu$NCE cross section on bound neutrons (per bound neutron). The efficiency corrections should be applied to the  predicted cross sections of the individual processes in order to compare with the MiniBooNE $\bar \nu$NCE scattering cross section result.

\bibliographystyle{apsrev_nourl}

\bibliography{references}

\section{Tables}
Here we tabulate the results presented in this paper. Table~\ref{tab:xs} lists the $\bar \nu$NCE differential cross section, the  $\bar \nu$NCE-like background (as shown in Fig.~\ref{fig:xs}), and the correction coefficients, in bins of $Q^{2}_{QE}$. Table~\ref{tab:nce_ratio} quantifies the anti neutrino NCE to neutrino NCE scattering cross section ratio measurement shown in Fig.~\ref{fig:nce_ratio}. And Table~\ref{tab:ccqe_ratio} lists the $\bar \nu$NCE to $\bar \nu$CCQE differential cross section ratio measurement shown in Fig.~\ref{fig:ccqe_ratio}.

\newpage

\begin{table}[htbp]
\begin{ruledtabular}
\begin{tabular}{cc}
$Q^{2}_{QE}(\mbox{GeV}^2)\backslash$ distribution &     $\frac{\sigma^{\bar\nu}_{NCE}}{\sigma^{\bar \nu}_{CCQE}}$ \\
\hline\noalign{\smallskip}                  
0.100--0.150 & $0.245 \pm 0.013$   \\
0.150--0.200 & $0.248 \pm 0.014$   \\
0.200--0.250 & $0.249 \pm 0.017$   \\
0.250--0.300 & $0.264 \pm 0.020$   \\
0.300--0.350 & $0.275 \pm 0.024$   \\
0.350--0.400 & $0.272 \pm 0.027$   \\
0.400--0.450 & $0.277 \pm 0.033$   \\
0.450--0.500 & $0.275 \pm 0.037$   \\
0.500--0.600 & $0.269 \pm 0.045$   \\
0.600--0.700 & $0.281 \pm 0.059$   \\
0.700--0.800 & $0.284 \pm 0.072$   \\
0.800--1.000 & $0.294 \pm 0.086$   \\
1.000--1.200 & $0.391 \pm 0.139$   \\
1.200--1.500 & $0.535 \pm 0.276$    \\
\end{tabular}
\end{ruledtabular}
\caption{MiniBooNE measured $\bar \nu$NCE/$\bar \nu$ CCQE cross section ratio as a function of $Q^2_{QE}=2m_N\sum_i T_i$.}
\label{tab:ccqe_ratio}
\end{table}

\begin{table}[htbp]
\begin{ruledtabular}
\begin{tabular}{cc}
$Q^{2}_{QE}(\mbox{GeV}^2)\backslash$ distribution &     $\frac{\sigma^{\bar\nu}_{NCE}}{\sigma^{\nu}_{NCE}}$
 \\
\hline\noalign{\smallskip}     
0.067--0.135 & $0.555\pm 0.0371$   \\
0.135--0.202 & $0.473\pm 0.0304$   \\
0.202--0.270 & $0.393\pm 0.0257$   \\
0.270--0.337 & $0.344\pm 0.0230$   \\
0.337--0.405 & $0.300\pm 0.0232$   \\
0.405--0.472 & $0.265\pm 0.0210$   \\
0.472--0.540 & $0.228\pm 0.0183$   \\
0.540--0.608 & $0.202\pm 0.0189$   \\
0.608--0.675 & $0.184\pm 0.0193$   \\
0.675--0.743 & $0.170\pm 0.0210$   \\
0.743--0.810 & $0.160\pm 0.0225$   \\
0.810--0.878 & $0.155\pm 0.0254$   \\
0.878--0.945 & $0.148\pm 0.0269$   \\
0.945--1.013 & $0.149\pm 0.0222$   \\
1.013--1.080 & $0.159\pm 0.0181$  \\ 
1.080--1.148 & $0.157\pm 0.0267$  \\ 
1.148--1.216 & $0.151\pm 0.0283$  \\ 
1.216--1.283 & $0.144\pm 0.0187$  \\ 
1.283--1.351 & $0.138\pm 0.0163$  \\ 
1.351--1.418 & $0.139\pm 0.0169$  \\ 
1.418--1.486 & $0.132\pm 0.0159$  \\ 
1.486--1.553 & $0.132\pm 0.0180$  \\ 
1.553--1.621 & $0.141\pm 0.0283$  \\ 
1.621--1.689 & $0.136\pm 0.0317$  \\
\end{tabular}
\end{ruledtabular}
\caption{MiniBooNE $\bar \nu$NCE/$\nu$NCE scattering cross section ratio measured as a function of $Q^2_{QE}=2m_N\sum_i T_i$.}
\label{tab:nce_ratio}
\end{table}

\begin{table*}
\begin{ruledtabular}
\begin{tabular}{cccccc}
$Q^{2}_{QE}(\mbox{GeV}^2)\backslash$ distribution & $\bar \nu$NCE cross section, (cm$^2$/GeV$^2$)     &$\bar \nu$NCE-like background, (cm$^2$/GeV$^2$) & $C_{\bar \nu p, H}$& $C_{\bar \nu p, C}$ & $C_{\bar \nu n, C}$  
\\
\hline\noalign{\smallskip}     
0.033--0.100 & $(1.009 \pm 0.068)\times 10^{-39}$  & $1.442\times 10^{-41}$ &0.484 &1.017 &1.229 \\
0.100--0.168 & $(1.740 \pm 0.106)\times 10^{-39}$  & $3.995\times 10^{-41}$ &1.134 &1.128 &0.868 \\
0.168--0.235 & $(1.268 \pm 0.083)\times 10^{-39}$  & $4.068\times 10^{-41}$ &1.055 &1.050 &0.944 \\
0.235--0.303 & $(9.175 \pm 0.719)\times 10^{-40}$  & $4.280\times 10^{-41}$ &1.028 &1.018 &0.979 \\
0.303--0.370 & $(6.766 \pm 0.633)\times 10^{-40}$  & $5.350\times 10^{-41}$ &0.998 &0.997 &1.002 \\
0.370--0.438 & $(4.903 \pm 0.551)\times 10^{-40}$  & $6.904\times 10^{-41}$ &0.972 &0.984 &1.018 \\
0.438--0.506 & $(3.531 \pm 0.457)\times 10^{-40}$  & $8.734\times 10^{-41}$ &1.001 &0.981 &1.019 \\
0.506--0.573 & $(2.510 \pm 0.406)\times 10^{-40}$  & $1.033\times 10^{-40}$ &0.979 &0.983 &1.020 \\
0.573--0.641 & $(1.820 \pm 0.346)\times 10^{-40}$  & $1.120\times 10^{-40}$ &0.986 &0.980 &1.022 \\
0.641--0.708 & $(1.337 \pm 0.294)\times 10^{-40}$  & $1.147\times 10^{-40}$ &0.981 &0.985 &1.018 \\
0.708--0.776 & $(9.926 \pm 2.511)\times 10^{-41}$  & $1.139\times 10^{-40}$ &0.997 &0.985 &1.016 \\
0.776--0.844 & $(7.597 \pm 2.117)\times 10^{-41}$  & $1.081\times 10^{-40}$ &0.969 &0.980 &1.025 \\
0.844--0.911 & $(5.853 \pm 1.828)\times 10^{-41}$  & $9.813\times 10^{-41}$ &0.966 &0.969 &1.038 \\
0.911--0.979 & $(4.733 \pm 1.481)\times 10^{-41}$ & $8.883\times 10^{-41}$ &0.936 &0.973 &1.036 \\
0.979--1.046 & $(4.133 \pm 1.051)\times 10^{-41}$ & $8.280\times 10^{-41}$ &0.979 &0.983 &1.022 \\
1.046--1.114 & $(3.516 \pm 0.912)\times 10^{-41}$ & $7.394\times 10^{-41}$ &0.977 &0.980 &1.025 \\
1.114--1.181 & $(3.010 \pm 0.923)\times 10^{-41}$ & $6.644\times 10^{-41}$ &0.904 &0.976 &1.035 \\
1.181--1.249 & $(2.367 \pm 0.637)\times 10^{-41}$ & $5.519\times 10^{-41}$ &0.736 &0.893 &1.152 \\
1.249--1.317 & $(1.885 \pm 0.454)\times 10^{-41}$ & $4.384\times 10^{-41}$ &0.685 &0.865 &1.192 \\
1.317--1.384 & $(1.531 \pm 0.375)\times 10^{-41}$ & $3.640\times 10^{-41}$ &0.464 &0.867 &1.202 \\
1.384--1.452 & $(1.282 \pm 0.285)\times 10^{-41}$ & $3.055\times 10^{-41}$ &0.386 &0.851 &1.241 \\
1.452--1.519 & $(1.037 \pm 0.268)\times 10^{-41}$ & $2.576\times 10^{-41}$ &0.323 &0.811 &1.286 \\
1.519--1.587 & $(8.989 \pm 2.698)\times 10^{-42}$ & $2.160\times 10^{-41}$ &0.526 &0.840 &1.233 \\
1.587--1.655 & $(8.142 \pm 2.982)\times 10^{-42}$ & $1.958\times 10^{-41}$ &0.240 &0.854 &1.222 \\
\end{tabular}
\end{ruledtabular}
\caption{
MiniBooNE measured $\bar \nu$NCE differential cross-section, predicted $\bar \nu$NCE-like background, and predicted correction coefficients for the three different  $\bar \nu$NCE scattering contributions as a function of $Q^2_{QE}=2m_N\sum_i T_i$.}
\label{tab:xs}
\end{table*}


\end{document}